\documentstyle[epsf]{EuroPhys}
\input EuroMacr
\begin{document}
\euro{}{}{}{}
\Date{}
\shorttitle{E. McCann and K. Richter:
From clean to diffusive mesoscopic systems}
\title{
From clean to diffusive mesoscopic systems: \\
A semiclassical approach to the magnetic susceptibility}
\author{Edward McCann and Klaus Richter}
\institute{Max-Planck-Institut f\"{u}r Physik komplexer Systeme,
N\"othnitzer Str. 38, 01187~Dresden, Germany}
\rec{}{}
\pacs{
\Pacs{03}{65.Sq}{}
\Pacs{05}{45.$+$b}{}
\Pacs{73}{20.Dx}{}}
\maketitle
%

\begin{abstract}
We study disorder-induced spectral correlations and 
their effect on the magnetic 
susceptibility of mesoscopic quantum systems in the non-diffusive regime.
By combining a diagrammatic perturbative approach with semiclassical techniques
we perform impurity averaging for non-translational invariant systems. 
This allows us to study the crossover from clean to diffusive systems.
As an application we consider the susceptibility of non-interacting
electrons in a ballistic microstructure in the presence
of weak disorder.
 We present numerical results for a square billiard and 
approximate analytic results for generic chaotic geometries.
 We show that for the elastic mean free path $\ell$ larger
than the system size, there are two distinct regimes of behaviour depending 
on the relative magnitudes of $\ell$ and an inelastic scattering length.
\end{abstract}

Phase-coherent, disordered conductors where the electron motion 
equals a random walk between impurities have traditionally been of 
interest in mesoscopic physics \cite{AltLeeWebb:91}. 
Random walks occur in the diffusive regime where the 
elastic mean free path $\ell$ 
is much smaller than the system size $L$. 
On the other hand, the development of 
high-mobility semiconductor heterostructures, combined with 
advanced lithographic 
techniques, have allowed the confinement of electrons to two-dimensional 
microstructures of controllable, non-random geometry. They have been
coined  ``ballistic'' since $\ell > L$. 
Nevertheless, residual impurity scattering
is nearly unavoidable even in these systems and it became clear 
that disorder can be 
strong enough to mix energy levels and effect the two-level correlation 
function, $K(\varepsilon_1,\varepsilon_2)$,  
even in the ``ballistic'' regime \cite{S+I:86,A+G:95}.
There it is necessary to 
consider both disorder averaging, $\left< \ldots \right>_d$, 
and size (or energy) averaging, $\left< \ldots \right>_L$. 
After such averaging the two-level correlation function may be 
divided into two separate terms \cite{A+G:95}, 
$\left<\, K^d(\varepsilon_1,\varepsilon_2 ) \,\right>_L =
\langle \langle \nu(\varepsilon_1) \nu(\varepsilon_2) \rangle_d \rangle_L
- \langle \langle \nu(\varepsilon_1) \rangle_d 
\langle \nu(\varepsilon_2) \rangle_d \rangle_L$
and 
$K^L(\varepsilon_1,\varepsilon_2) =
\langle \langle \nu(\varepsilon_1) \rangle_d 
\langle \nu(\varepsilon_2) \rangle_d \rangle_L
- \bar{\nu}^2$, 
where $\nu$ denotes the single particle density of states 
and $\bar{\nu} = \langle \langle \nu(\varepsilon) \rangle_d \rangle_L$
its mean part.
 $K^d$ is a measure of 
{\em disorder-induced} correlations of $\nu$, while $K^L$ is given by
{\em size-induced} correlations.

The orbital magnetism of isolated mesoscopic systems has been the 
subject of much theoretical interest, in particular because it is sensitive
to spectral correlations: For a system with a fixed number of particles 
it is necessary to consider averaging under canonical conditions 
\cite{Che:88,B+M:89,Alt:91} resulting in a large
contribution to the average magnetism. The corresponding 
susceptibility is given by \cite{Alt:91}
$\left<\chi (H)\right> = - \left(\Delta /2\right) 
\partial^2/\partial H^2
 \left<\delta N^2 (\mu ;H)\right>.$
Here $H$ is the magnetic field, $\Delta$ is the mean level spacing and 
$\left<\delta N^2 (\mu ;H)\right>$ is the variance in the
number of energy levels within an energy interval of width 
equal to the chemical potential, $\mu$.
 This variance is related  to
$K(\varepsilon_1,\varepsilon_2 ;H)$ by integration of the
level energies $\varepsilon_1$, $\varepsilon_2$ over the energy interval.
 In the following we will label the contributions to 
the susceptibility, corresponding to $\left<\, K^d \,\right>_L$ and $K^L$, as
$\left<\chi^d (H)\right>$ and $\left<\chi^L (H)\right>$, 
respectively, so that
$\left<\chi (H)\right> = \left<\chi^d (H)\right>
+ \left<\chi^L (H)\right>.$

For real systems, besides $\ell$, 
there are additional relevant lengthscales at which 
inelastic scattering ($L_{\phi}$) or temperature smearing ($L_{T}$) 
produce a damping of propagation.
For clarity we refer to such a lengthscale as $L_{\phi}$, although we assume 
that  
similar general arguments will hold for finite $L_{T}$. For ballistic motion, 
$L_\phi$
is related to the level broadening $\gamma$ by 
\begin{equation}
\frac{L_\phi}{L} = \frac{k_F L}{2\pi} \frac{\Delta}{\gamma} 
\label{Lphi}
\end{equation}
with $k_F$ being the Fermi momentum.
It  divides the ``ballistic'' regime into two sub-regimes.
In the first, $L , L_{\phi} < \ell$, the particle motion is 
nearly ballistic since damping due to inelastic scattering 
typically occurs before impurity scattering; for the remainder of this
paper we refer to this regime as {\it inelastic}.
 In the second, $L < \ell < L_{\phi}$, a particle may scatter many times 
off impurities before scattering inelastically and we refer to 
this regime as {\it elastic}.

In this paper we employ a systematic semiclassical approach to calculate 
the contribution of orbits {\em of all lengths} in the elastic and 
inelastic regimes
to weak field $\left<\chi^d (H)\right>$ for microstructures 
with white noise disorder \cite{Agam:96}.
 Energy and impurity averaging are performed within a diagrammatic 
perturbation approach applicable to the ``ballistic'' regime.
 In contrast to many techniques valid for bulk disordered systems, 
$\ell <L$, we do not assume translational invariance.
 Instead, we use an approach related to the ``method of trajectories'' 
devised for thin superconducting films \cite{G+T:64} 
and we write Green functions semiclassically in terms of classical 
paths which include the effect of boundary scattering.
 Note that similar \cite{B+H:88}, and alternative \cite{D+K:84}, 
methods have been used to consider weak localization in 
thin films in a parallel magnetic field.
Our approach allows us to study the complete crossover from diffusive  
to clean systems for arbitrary values of $L_{\phi}$ 
(smaller than $v_F t_H$, where $t_H$ is the Heisenberg time).
We present results for ballistic systems with both chaotic and 
integrable dynamics in the clean limit.
 As an example of an integrable geometry, we treat the case of the 
square billiard.
 Experiments \cite{Lev:93} on the orbital magnetism of ensembles of 
squares were performed in the
``ballistic'' (inelastic) regime, motivating theoretical studies of 
the susceptibility 
for $L < \ell$ \cite{Gef:94,Ric:96}.
\section{Semiclassical diagrammatic approach}
We begin by presenting some more details concerning our semiclassical 
evaluation of the disorder correlation function, $K^d$, and 
corresponding susceptibility $\left<\chi^d (H)\right>$.
 We consider non-interacting electrons in a 
weak, perpendicular magnetic field.
 In terms of retarded and advanced single particle Green functions, 
${\cal G}^{+(-)}({\bf  r}_1 , {\bf  r}_2 ; \varepsilon ;H)$, $K^d$ 
may be written as 
$K^d(\varepsilon_1,\varepsilon_2 ;H) \approx
\left( \Delta^2/2\pi^2\right) {\cal R} \left<\!\left<
{\rm tr} \,{\cal G}^{+}(\varepsilon_1 ;H) {\rm tr} \,
{\cal G}^{-}(\varepsilon_2 ;H)
\right>\!\right>_d$, 
where $\left<\!\left< \ldots \right>\!\right>_d$ implies the inclusion 
of connected diagrams only.  Using a diagrammatic approach
introduced by Altland and Gefen \cite{A+G:95}
the field sensitive part of $K^d$ can be expressed as a sum including
Cooperon type diagrams ${\cal S}_{n}^{(C)}$ defined by
\begin{equation}
{\cal S}_{n}^{(C)}(\omega ;H) = {\rm Tr} \left[\zeta^{(C)}\right]^n =
\int \prod_{j=1}^{n} d^dr_j\prod_{m=1}^{n}
\zeta^{(C)} ({\bf r}_m,{\bf r}_{m+1};\omega ;H) \quad ; \quad 
{\bf r}_{n+1} \equiv {\bf r}_1 \; .
\label{sn}
\end{equation}
Here
$\zeta^{(C)} ({\bf  r}_1 , {\bf  r}_2 ;\omega ;H)
= \left( 1/2\pi\bar{\nu}\tau \right)
G^{+}({\bf  r}_1 , {\bf  r}_2 ; \varepsilon_1;H )
G^{-}({\bf  r}_1 , {\bf  r}_2 ; \varepsilon_2;H )$, 
$G^{+(-)} = \left< {\cal G}^{+(-)} \right>_d$ is the disorder 
averaged single particle Green function, 
$\omega = \varepsilon_1 - \varepsilon_2$, and $\tau = \ell/v_F$.

Semiclassically, $G^+({\bf r}_1, {\bf r}_2)$ can be expressed as a 
sum over classical trajectories $t$ between ${\bf  r}_1$ and ${\bf  r}_2$
\cite{Ric:96}, 
\begin{equation}
 G^{+}({\bf  r}_1 , {\bf  r}_2) \simeq 
\sum_t D_t ({\bf  r}_1 , {\bf  r}_2)
 \exp{\left[\frac{i}{\hbar} S_t({\bf r}_1,{\bf r}_2)
-\frac{L_t({\bf  r}_1 , {\bf  r}_2)}{2\ell}\right]} \; .
\end{equation}
The prefactor $D_t$ includes the classical phase space density, 
$S_t$ stands for the classical action along an orbit $t$ 
(in the absence of disorder)
including the Maslov index, and 
$L_t$ is the orbit length.
$\zeta^{(C)}({\bf  r}_1,{\bf  r}_2;\omega ;H)$ is then  given 
in terms of pairs of classical paths which explicitly include 
the effect of boundary scattering.
However most pairs (of different paths) produce oscillating 
contributions which 
we assume to vanish after energy or size averaging\footnote{
In a ballistic system additional
oscillatory terms will however remain upon pure disorder average 
for fixed size \cite{A+G:95,M+R:98}.}.
 The main contribution to the field sensitive part of 
$\left<\, K^d \,\right>_L$ arises from diagonal terms 
(otherwise known as the Cooperon channel) obtained by pairing paths with 
their time reverse.
Assuming that the magnetic field affects 
the phase of the particles but not their trajectories we can write
$\zeta^{(C)} ({\bf  r}_1, {\bf  r}_2 ; \omega ;H)
= \sum_{t:{{\bf  r}_1} \rightarrow {{\bf r}_2}}
\tilde\zeta_t^{(C)} ({\bf  r}_1, {\bf  r}_2 ; \omega ;H)$
where
\begin{equation}
\tilde\zeta_t^{(C)} ({\bf  r}_1 , {\bf  r}_2 ; \omega ;H)
\simeq 
\frac{v_F |D_t|^2}{2\pi\bar{\nu} \ell} \exp \left[ 
- \frac{L_t}{L_\phi} - \frac{L_t}{\ell}  + i\omega T_t
+ i \frac{4\pi}{\varphi_0}
\int_{{\bf  r}_1}^{{\bf  r}_2} {\bf A.dr} \right] \; .
\label{zsc}
\end{equation}
Here, $T_t$ is the period of the trajectory, $\bf A$ is the vector potential, 
$\varphi_0 = hc/e$, and the level broadening was introduced via
$\omega \rightarrow \omega + i\gamma$. 
Eq.~(\ref{zsc}) depends, besides $\ell$, only on the system
without disorder and holds for both integrable and chaotic geometries.

The disorder induced contribution to the average susceptibility is given by
($\varphi = H L^2/ \varphi_0$): 
\begin{equation}
\frac{\left< \chi^d (\varphi) \right>}{\left| \chi_L \right|} \approx
- \frac{6}{\pi^2}  \frac{\partial^2}{\partial \varphi^2} \
\sum_{n=1}^{\infty} \frac{1}{n} {\cal S}_{n}^{(C)} (0 ;\varphi) \; ,
\label{stos}
\end{equation}
where the bulk Landau susceptibility is 
$\chi_L = -e^2/24\pi mc^2$ for spinless electrons.
In Eq.~(\ref{stos}) the ${\cal S}_n^{(C)}$ are now assumed to contain 
diagonal terms only and their contributions 
(Eq.~(\ref{sn})) 
can be calculated by diagonalising $\zeta^{(C)}$ which in general cannot
be done analytically. However we can use the fact that all the 
variations of $\zeta^{(C)}$ occur on classical lengthscales; rapid 
oscillations on the scale of $\lambda_F$ cancel out.
It is thus possible to discretise the ``classical'' operator
$\zeta^{(C)}$ on a lattice in 
space with grid size greater than $\lambda_F$. By summing over 
trajectories between lattice cells one can compute
 $\zeta^{(C)}$, and thereby  $\left< \chi^d (H) \right>$,
efficiently by numerical means\cite{M+R:98}\footnote{
A similar calculation has been applied to a different
``classical'' operator, describing interaction effects in 
ballistic quantum dots, in Ref.~\cite{Ull:97}.}.

The propagator $\zeta^{(C)}$ (above Eq.~(\ref{zsc})) is made up 
of a summation over all diagonal pairs of paths (including boundary 
scattering) between any two given impurities situated at 
${\bf  r}_1$ and  ${\bf  r}_2$.
 On taking the trace over $n$ propagators $\zeta^{(C)}$, one sees
that the field sensitive part of $S_n$, Eq.~(\ref{sn}), 
consists of a summation over flux-enclosing closed pairs of paths 
(in position space) involving $n$ impurities and an arbitrary number of 
boundary scattering events.
 However this summation does not include closed pairs of paths 
which follow periodic 
orbits of the corresponding clean system.
 Such paths involve zero momentum transfer between 
the Green functions at the impurity positions; they actually represent 
disconnected diagrams which are included in $\left<\chi^L (H)\right>$
and must not be counted again in the determination of 
$\left<\chi^d (H)\right>$.
 It is the presence of these periodic orbits in the determination of
$\left<\chi^L (H)\right>$
\footnote{A similar sensitivity 
with respect to the system geometry occurs in interacting 
ballistic quantum dots due to the presence of off-diagonal periodic orbit
terms \cite{Ull:97}.}
 that leads to strong sensitivity 
with respect to the system geometry \cite{Ull:95,Ric:96a} (see below).

In the following we apply the above formalism to the case of an ensemble
of disordered square billiards. 
Gefen, Braun, and Montambaux (GBM) \cite{Gef:94} considered 
the contribution of trajectories longer than $\ell$ to 
$\left<\chi^d (H)\right>$ in an approximate way,
while Richter, Ullmo, and Jalabert (RUJ) \cite{Ric:96} 
calculated $\left<\chi^L (H)\right>$ for a square by assuming that 
the disorder perturbs the phase, but not the trajectory, of 
semiclassical paths of the corresponding clean geometry. 
We first compute the ${\cal S}_{n}^{(C)}$ (Eq.~(\ref{sn})) for
the square geometry by employing the extended zone scheme\cite{Ric:96}
to write  $\zeta^{(C)} ({\bf  r}_1 , {\bf  r}_2 ; \omega ;H)$
as a sum of propagators along straight line paths.
We then perform a complete calculation of 
$\left<\chi^d (H)\right>$, Eq.~(\ref{stos}), 
in the elastic and inelastic regimes and 
compare  the results with those by GBM and RUJ.
%
%
%
\begin{figure}
\centerline{\epsfxsize=0.5\hsize \epsffile{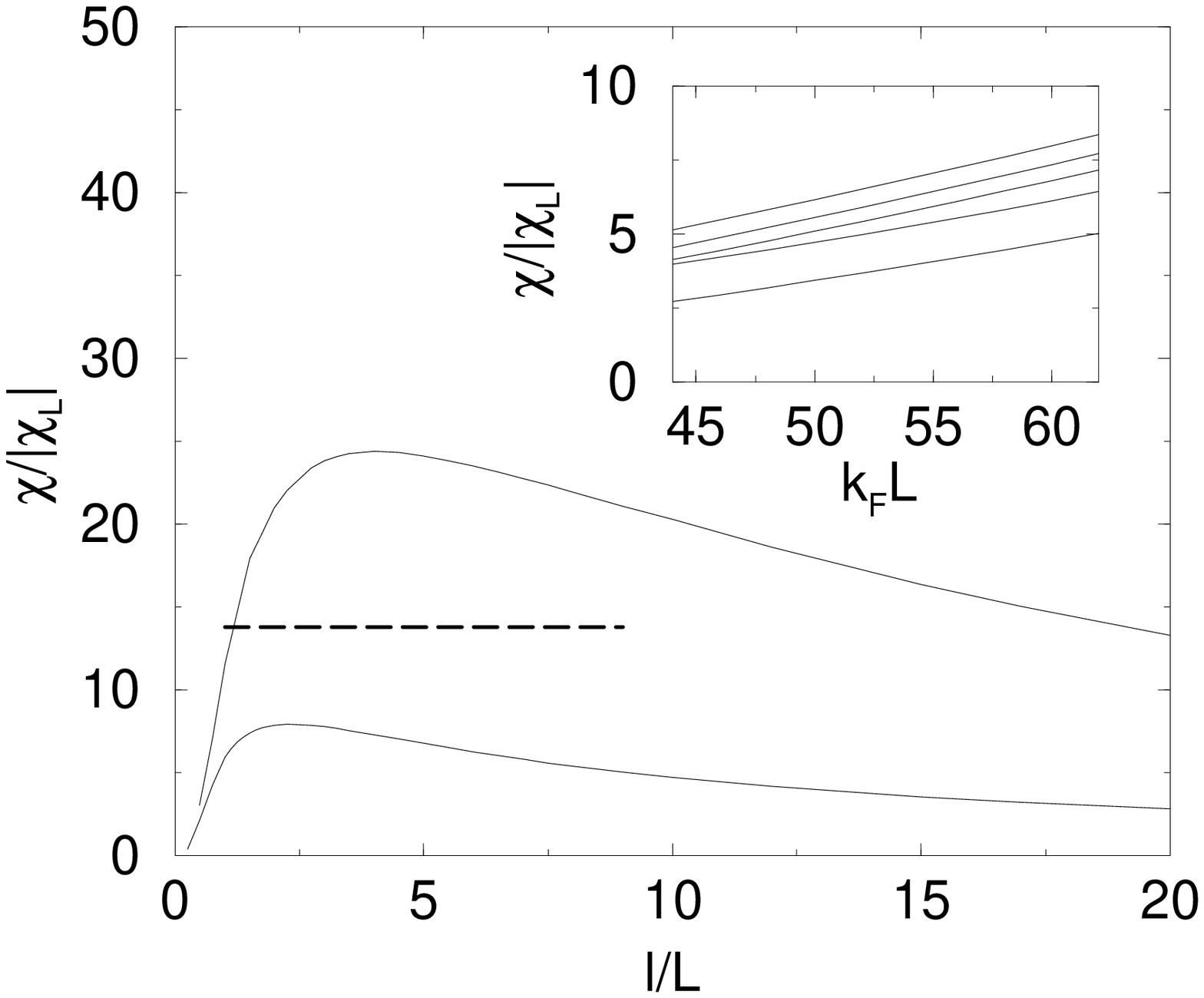}
\epsfxsize=0.52\hsize \epsffile{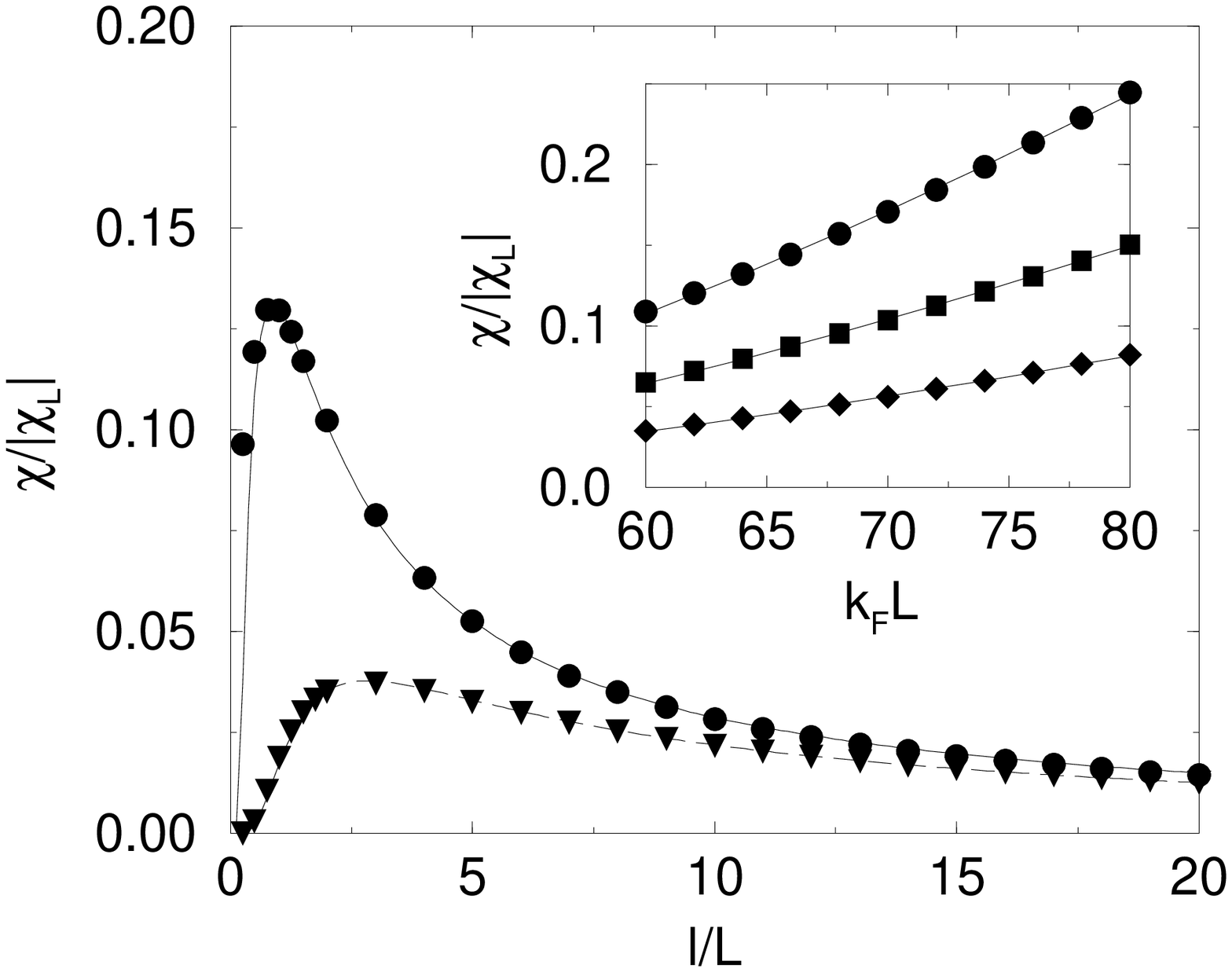}}
\label{elastic}
\caption{Disorder-induced average susceptibility $\left<\chi^d (0)\right>$ for
a square geometry in the elastic regime ($L < \ell < L_{\phi}$) 
as a function of the elastic mean free path $\ell$ for $k_FL=60$ 
and two strengths of inelastic scattering, 
$\gamma /\Delta = 1$ (lower), which corresponds to 
$L_{\phi} /L \approx 9.5$, and $\gamma /\Delta = 0.392$ (upper).
The dashed horizontal line indicates the result by GBM
for $\gamma /\Delta = 1$.
The inset shows $\left<\chi^d (0)\right>$ as a function of $k_FL$ 
for $\gamma /\Delta = 1$.
 From the top, the 5 curves are for values of 
$\ell /L = 2$, $4$, $5$, $1$ and $10$.
}
\caption{$\left<\chi^d (0)\right>$ in 
the inelastic regime ($L,  L_{\phi} < \ell$) 
as a function of $\ell$ for $k_FL=60$ and $\gamma /\Delta = 10$ 
which corresponds to $L_{\phi} /L \approx 0.95$.
 Circles are our numerical results and the triangles are 
numerical results including only the contribution of $S_1$.
 The solid and dashed lines are both fits (see main text).
 The inset shows $\left<\chi^d (0)\right>$ as a function 
of $k_FL$ for $\gamma /\Delta = 10$.  From the top, the curves are for 
values of $\ell /L = 2$, $4$ and $8$.
 The symbols correspond to our numerical results and the solid lines to a fit.
}
\end{figure}

\section{Elastic regime: $L < \ell < L_\phi$}
Fig.~1 shows $\left<\chi^d (0)\right>$ as a function 
of $\ell$ for a typical experimental value of $k_FL=60$. 
The lower curve is for $\gamma /\Delta = 1$  
 (i.e.\  $L_{\phi} /L \approx 9.5$ at $k_FL=60$)
and the upper is for $\gamma /\Delta = 0.392$ \cite{note2}.
 Note that in the diffusive regime, $\ell <L$, there is linear increase 
with $\ell$ in agreement with Ref.~\cite{Oh:91}.
For $L < \ell < L_{\phi}$, we find a weak dependence of 
$\left<\chi^d (0)\right>$ on $\ell$. Our result
is on the whole close to the prediction by GBM
\cite{Gef:94} who found a paramagnetic $\ell$-independent contribution, 
$ \left< \chi^d (0) \right>/\left| \chi_L \right| \approx 
0.23 \ k_FL (\Delta/\gamma) $,
shown as the dashed horizontal line in Fig.~1 for $\gamma /\Delta = 1$.
As $\ell$ increases further, $\left<\chi^d (0)\right>$ decreases; 
we discuss this behaviour in more detail later when considering the
inelastic regime.

The inset of Fig.~1 shows $\left<\chi^d (0)\right>$ as a function of 
$k_FL$ for $\gamma /\Delta = 1$.
 From the top, the five curves are for values of 
$\ell /L = 2$, $4$, $5$, $1$ and $10$.
 For all disorder strengths, $\left<\chi^d (0)\right>$ 
is paramagnetic and it increases linearly with $k_FL$.
 For $L < \ell <  L_{\phi}$ the gradient of the curves 
is approximately independent of $\ell$ and we find it to be 
$\approx 0.18 \Delta /\gamma$.
 However there is a $k_FL$ independent offset to the curves 
which is $\ell$ dependent and not described by GBM.
\section{Inelastic regime: $L, L_\phi < \ell$}
Fig.~2  shows $\left<\chi^d (0)\right>$ as a function 
of $\ell$ for $k_FL=60$ and $\gamma /\Delta = 10$ 
(i.e.\ $L_{\phi} /L \approx 0.95$).
Circular points correspond to our full numerical results, while
triangular points represent the contribution of $S_1$ only 
(minus the disconnected part) in Eq.~(\ref{stos}).
The solid line is the equation
$\left<\chi^d (0)\right> = (a_0L/\ell)   
 \exp\left( -a_1L/\ell\right)$
where $a_0$ and $a_1$ are fitting parameters. 
The dashed line displays an equation of the same type
but with $a_1 = 2\sqrt{2}$ and $a_0$ as the only fitting parameter.
For $\ell \gg L$ the susceptibility is dominated by the contribution
with the lowest number of impurity scatterings $S_1$.
 As $\ell /L$ is reduced, progressively more 
terms in the summation of Eq.~(\ref{stos}) become relevant,
and there is good agreement with the solid line fit for 
$\ell \geq L$.

The inset of Fig.~2 shows $\left<\chi^d (0)\right>$ as a function of 
$k_FL$ for $\gamma /\Delta = 10$.  From the top, the three curves are 
for values of $\ell /L = 2$, $4$ and $8$.
 The symbols correspond to our results and the solid lines to
$\left<\chi^d (0)\right> = b_0\exp
\left[ -b_1(\gamma /\Delta )/k_FL \right]$
where $b_0$ and $b_1$ are fitting parameters.
%
%
\begin{figure}
\centerline{\epsfxsize=0.48\hsize \epsffile{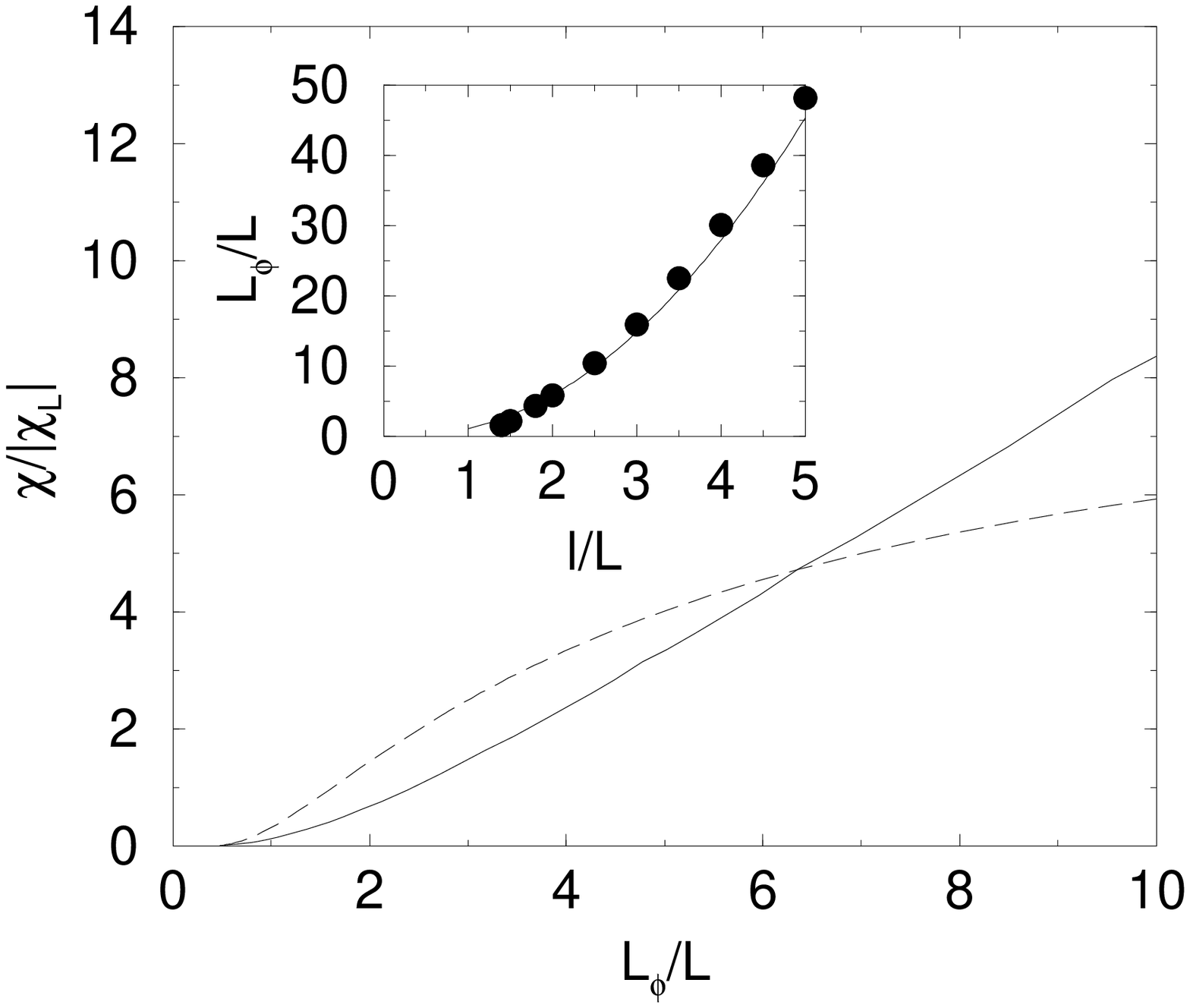}
\epsfxsize=0.48\hsize \epsffile{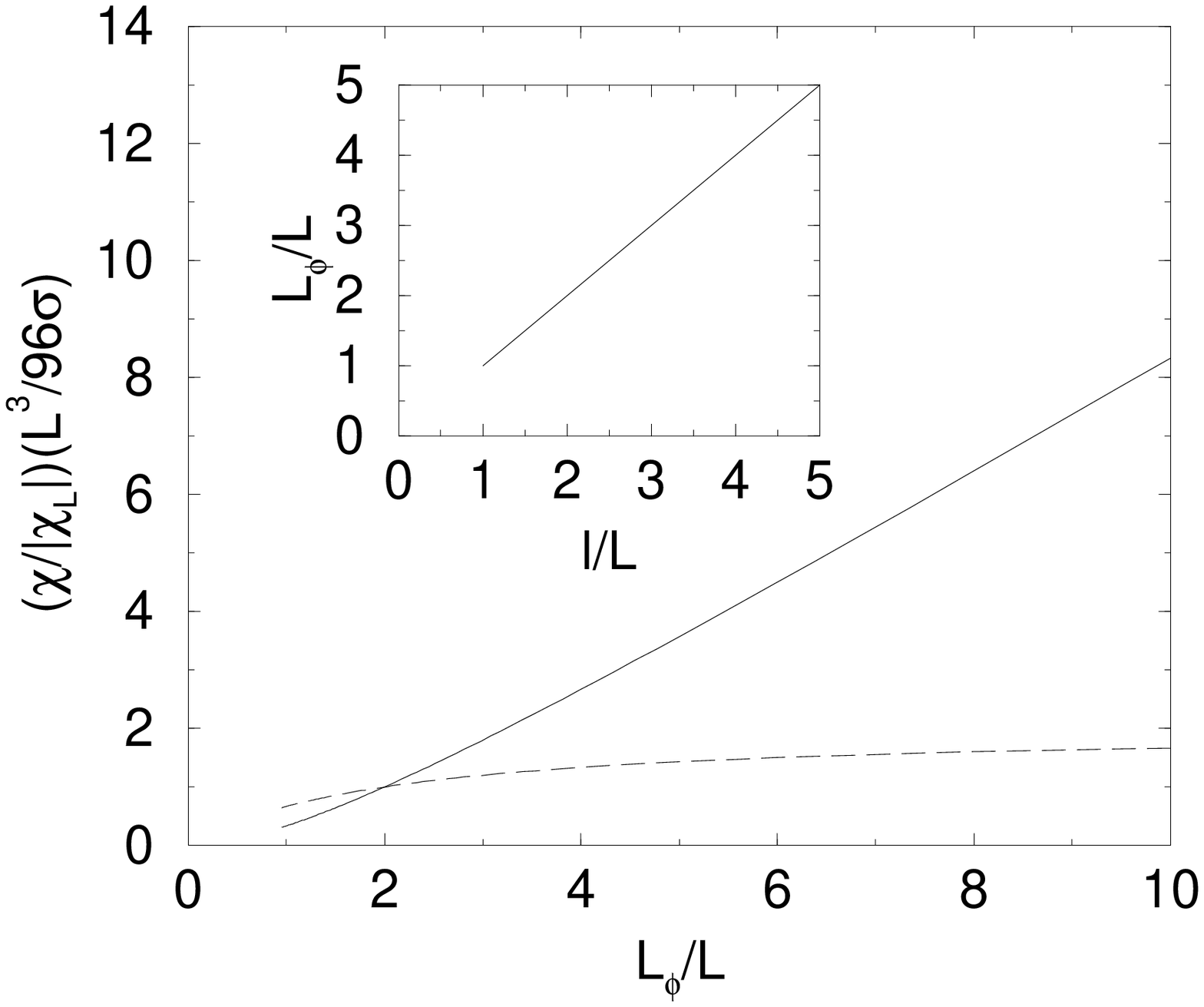}}
\label{compare}
\caption{Comparison of $\left< \chi^d (0) \right>$ 
and $\left< \chi^L (0) \right>$ for the square.
 The solid line shows our numerical results for 
$\left<\chi^d (0)\right>$ and 
the dashed line is an analytical expression for 
$\left<\chi^L (0)\right>$ (see main text) as a function of 
$L_\phi/ L$ for $k_FL=60$ and $\ell /L =2$.
Inset: value of $L_\phi /L$ at which the two 
contributions are equal as a function of $\ell / L$.
The solid line is the analytical estimate Eq.~(\protect\ref{cgam})
and the circles are 
obtained by comparing our numerical results for 
$\left<\chi^d (0)\right>$ with  
the analytic expression for $\left<\chi^L (0)\right>$.}
\caption{Comparison of the (normalized) 
semiclassical estimates for $\left< \chi^d (0) \right>$ 
(solid line) and $\left< \chi^L (0) \right>$ (dashed) 
for a generic chaotic geometry for $k_F L = 60$ (see main text).
Inset: the (straight) line shows the values in the $(\ell,L_\phi)$-plane
where both contributions are equal.
}
\end{figure}


%
%
\section{Comparison with the contribution of clean correlations}
We compare the magnitude of $\left<\chi^d (0)\right>$ with 
that of $\left<\chi^L (0)\right>$ for squares.
 It has been shown \cite{Ull:95,Ric:96a} that the low field 
susceptibility of an ensemble of {\em clean} squares is dominated by the 
shortest flux enclosing periodic orbits of length $L_t = 2 \sqrt{2}L$ 
and their repetitions over a 
broad range of temperature (and thus inelastic scattering strengths).
For the ballistic white noise case considered here, 
the effect of disorder averaging
 on the susceptibility was described by an additional damping
$\exp ( - L_t /\ell )$ of the response of the clean system \cite{Ric:96}.
This result corresponds to $\left<\chi^L (0)\right>$ including 
the disorder damping $\exp ( - L_t /2\ell )$ of the 
single particle Green functions.

We use the results of
RUJ \cite{Ric:96,Ull:95} at zero temperature and introduce 
the level broadening $\gamma$ in the same way as for the disorder 
correlations above \cite{note1}.
 It is then possible to sum the 
contribution of all repetitions of the fundamental orbit 
explicitly which gives 
$
\left< \chi^L (0) \right>/\left| \chi_L \right| \simeq
(\sqrt{2}/5\pi) k_FL/\sinh^2[\sqrt{2}(L/\ell + L/L_\phi)]
$.
Fig.~3 shows this expression for $\left<\chi^L (0)\right>$ (dashed line) and
our numerical results for $\left<\chi^d (0)\right>$ (solid line) 
as a function of $L_\phi / L $ for $k_FL=60$ and $\ell /L =2$.
 Although $\left<\chi^d (0)\right> < \left<\chi^L (0)\right>$ 
for small $L_\phi / L$ and vice versa for 
large $L_\phi / L$, it is clear that both contributions are 
relevant over a broad range of $L_\phi / L$.
 We make an estimate for the value of $L_\phi / L$ at 
which the contributions of $\left<\chi^L (0)\right>$ 
and $\left<\chi^d (0)\right>$ are equal by comparing the above
analytic approximation with that given by GBM.
We find for $\ell >L$ that $\left<\chi^d (H)\right>$ is larger 
than $\left<\chi^L (H)\right>$ for 
$L_\phi$ greater than a crossover value, $L^{*}_\phi$, given by
\begin{equation}
\frac{L^{*}_\phi}{L} \sim \frac{(k_F L)^2}{2\pi^2}\left[
\sqrt{2}k_F L \left( \frac{L}{\ell} \right)^2
+ 8\pi^2\left( \frac{L}{\ell} \right)^3\right]^{-1} \; .
\label{cgam}
\end{equation}
The inset of Fig.~3 shows this estimate (solid line) 
compared to points (circles) obtained by comparing our numerical 
results for $\left<\chi^d (0)\right>$ with  
the analytic expression for $\left<\chi^L (0)\right>$.

The experiment\cite{Lev:93} on the orbital magnetism of ensembles of 
squares had estimated values for the elastic mean free path
of $\ell /L \sim 1-2$, for the phase-coherence length 
of $\sim (3-10) L $ and for the thermal cutoff length
of $L_T /L \sim 2$.
Hence, the lengthscale $L_\phi$  (Eq.~(\ref{Lphi})) is determined by
the shorter length $L_T$. Fig.~3 shows that for these experimental
parameters (and for white noise disorder)
both the disorder and size-induced 
correlations are relevant, however the latter contribution is dominant.
The measured value of the susceptibility at {\em low}
temperature was $\chi(0) \sim 100 |\chi_L|$, 
with an uncertainty of about a factor of four.
After including a spin factor of 2,
 the combined contributions 
$\langle\chi^d\rangle$ and $\langle\chi^L\rangle$ 
calculated 
above, together with an interaction contribution of the same order
\cite{Ull:97}, are in broad agreement with the experimental result. 
We note, however, that a theoretical explanation of
the temperature dependence of the measured susceptibility 
is still lacking.

Experimental ballistic structures as those in Ref.~\cite{Lev:93}
are usually characterised by smooth disorder potentials.
The effect of smooth disorder on $\langle \chi^L \rangle$ has been analysed
in Ref.~\cite{Ric:96} showing that the reduction of the
clean contribution is less strong than for white noise disorder
and no longer exponential.
Smooth disorder effects can be incorporated
into the present calculation by introducing an angle-dependent
cross section for the impurity scattering between two 
trajectory segments.
\section{Chaotic geometries}
For systems with a generic chaotic, clean counterpart we obtain 
an analytical estimate for $\left<\chi^d (0)\right>$ in the elastic regime
after transforming the
sum over densities $|D_t|^2$ in Eq.~(\ref{zsc})
into probabilities $P({\bf r},{\bf r}';t|A)$ to
propagate classically from ${\bf r}$ to ${\bf r}'$ at time $t$ accumulating
an ``area'' $A$ \cite{M+R:98}.  Assuming a Gaussian ``area'' distribution 
with a variance $\sigma$, which is taken to be 
$\ell$ independent, we find
\begin{equation}
{\cal S}_{n}^{(C)}(\omega ;H) \approx \left\{1 +
\frac{8\pi^2 H^2 \ell \sigma}{\varphi_{0}^2} + (\gamma- i\omega)\tau
\right\}^{-n} \; .
\label{snch}
\end{equation}
Summation of the ${\cal S}_{n}^{(C)}$, Eq.~(\ref{stos}), leads to
\begin{equation}
\frac{\left<\chi^d(0) \right>}{|\chi_L|} 
\simeq \frac{96 \sigma L_\phi^2}{L^4(L_\phi+\ell)} \; .
\end{equation}
 $\left<\chi^d (0)\right>$ is shown as the full line in 
Fig.~4 which is rather similar
to the corresponding curve in Fig.~3 for the square geometry.
A corresponding approximation for 
$\langle \chi^L(0) \rangle$ \cite{Ric:96a} is shown
as the dashed line in Fig.~4.
Both can be shown to add up to an $\ell$-independent response
$\left<\chi(0)\right>/|\chi_L| \simeq 96 \sigma L_\phi /L^4$ 
for generic chaotic geometries.
\section{Conclusion}
Disorder-induced spectral correlations and 
their effect on the magnetic 
susceptibility in the non-diffusive regime $\ell > L$ were considered.
 We focused on the square billiard, whose corresponding clean
geometry is integrable, and showed that there are two distinct regimes of 
behaviour depending on the relative magnitudes of $\ell$ and $L_{\phi}$.
 This approach enabled us to study the complete crossover from diffusive  
to clean systems for arbitrary values of $\ell$ and $L_{\phi}$ 
(smaller than $v_F t_H$).
 Note that it may be possible to calculate the susceptibility for
values of $L_{\phi}$ and $\ell$
greater than $v_F t_H$ using a non-perturbative 
approach such as the ballistic $\sigma$ model \cite{M+K:95}.

\stars
We are grateful to Y.~Gefen, S.~Kettemann, D.~E.~Khmelnitskii, 
M.~Leadbeater, I.~V.~Lerner and P.~Walker for useful discussions.
 We thank the Isaac Newton Institute for Mathematical Sciences, Cambridge, 
where part of this research was performed.
%
%


\begin{thebibliography}{99}
\bibitem{note1} In Ref.~\cite{Ric:96} the orbits are damped by  
including temperature smearing rather than level broadening.
\end{thebibliography}
\end{document}